# A Taxonomy for Dynamic Honeypot Measures of Effectiveness

Jason M. Pittman, Kyle Hoffpauir, Nathan Markle, Cameron Meadows
*jpittman, khoffpau, nmarkle, cmeadows {@highpoint.edu}*
High Point University, High Point, NC

Abstract

Honeypots are computing systems used to capture unauthorized, often malicious, activity. While honeypots can take on a variety of forms, researchers agree the technology is useful for studying adversary behavior, tools, and techniques. Unfortunately, researchers also agree honeypots are difficult to implement and maintain. A lack of measures of effectiveness compounds the implementation issues specifically. In other words, existing research does not provide a set of measures to determine if a honeypot is effective in its implementation. This is problematic because an ineffective implementation may lead to poor performance, inadequate emulation of legitimate services, or even premature discovery by an adversary. Accordingly, we have developed a taxonomy for measures of effectiveness in dynamic honeypot implementations. Our aim is for these measures to be used to quantify a dynamic honeypot's effectiveness in fingerprinting its environment, capturing valid data from adversaries, deceiving adversaries, and intelligently monitoring itself and its surroundings.

1. Introduction

It is not a surprise that as the number of network-based computing services increases within globalized cyberspace, the breadth of cybersecurity attacks and frequency of breaches increase likewise. For example, there was a 56% increase in web-based attacks in 2018 [1]. Ransomware infections rose by 12% in the same period [1]. Overall, systems were attacked roughly every 40 seconds [2]. Perhaps these trends continue year after year because researchers and cybersecurity professionals have a broad understanding of what constitutes a specific type of attack but details of adversary behavior within a system are overshadowed by the need to simply defend.

With that said, one type of technology that can be used to study adversary interactions is honeypots. Honeypots attract adversaries by emulating operating systems, applications, and services with known vulnerabilities. Unfortunately, honeypots are extremely difficult to implement and maintain [4][5]. Dynamic honeypots are specifically problematic in this manner [6]. In fact, existing literature contains a plethora of suggestions as to how dynamic honeypots can be effectively deployed or managed. However, there is little quantitative validation of effectiveness in this regard which leaves professionals, researchers, and educators without the means to differentiate between implementation or management modalities.

Accordingly, this work presents a taxonomy for dynamic honeypot measures of effectiveness. The aim was to codify what *features* are common across dynamic honeypots into an actionable knowledge model. The significance of doing so rests in providing a quantitative means to determine whether a given dynamic honeypot is functioning within operational guidelines. However, before delving into how we developed our taxonomy of the presentation and details of the taxonomy, it will be helpful to discuss the background and motivation for this work.

## 2. Background

The relevant background for this work includes honeypot technology, measures of effectiveness, and taxonomies. We provide these discussions with the goal of contextualizing both the motivation for our study as well as the results. Accordingly, each background section presents a summary snapshot of existing research which we have curated with that goal in mind.

### 2.1 Honeypots

A honeypot is a computing system intended to attract adversaries and designed to be attacked [7]. Honeypots are qualitatively categorized with traits such as interaction level, deployment modes, and deployment categories to place them into four types being: shadow honeypots, honeynets, honeyfarms, and honeytokens [8]. Other recent research showcases various deceptive technologies and countermeasures to honeypot technology [9]. Building off this idea of deception, approaches were created to increase the deceptive nature of honeypot deployments through five methods [10]. Foremost amongst modern innovation in honeypots, researchers [11] have started to fuse machine learning and honeypot technology.

### 2.2 Measures of Effectiveness

A measure of effectiveness (MoE) can be described as a measure of performance which corresponds to a specific user requirement [12]. Thus, a MoE is typically used to assess whether a specific tool or service accomplishes its task. By defining the MoE for honeypots, we were able to expand an existing taxonomy [13] so that metrics could be created to assess and compare honeypot effectiveness. In the context of this study, such an ability to quantitatively assess and compare honeypots with our MoE satisfies the research question. For our purposes, we needed to determine user requirements for our honeypots. Specifically, we were able to determine that fingerprinting, data capture, deception, and intelligence are suitable MoE.

### 2.3 Taxonomies

For clarity, we take a taxonomy to be a classification of classes with values to which objects of the study are mapped based on the object's characteristics" [13]. There are two types of taxonomies that can be used in classification, being faceted and enumerative [13][14]. Importantly, a taxonomy utilizes a set of semantically cohesive categories that are combined as needed to create an expression of a concept [14]. For our purposes, we needed to utilize an enumerative taxonomy to incorporate our measures of effectiveness as cohesive and discrete categories. From this established taxonomy, we then created a hierarchical classification to categorize the honeypots [15].

### 3. Method

The motivation for this work is best expressed through a single research question: what, if any, existing features of dynamic honeypots can be used to model a set of measures of effectiveness. We developed this question based on the observation that existing literature acknowledged the problem of configuring and maintaining dynamic honeypots, and further existing literature asserts potential workarounds, there appeared to be no discussion as to how to measure whether a given honeypot was functioning as intended. Further, our preliminary work [16] was well received and thus prompted continued investigation. Therefore, the next logical step then was to construct a full, representative MoE model. The hope is that later work may transition our conceptual model into a practical tool.

3.1 Design
The first course of action to enable our work was to select a research design. Here, we considered grounded theory. However, we did not intend to produce theory or theoretical models. We also considered ontology. However, established conceptual frameworks for both honeypot and measures of effectiveness knowledge domains already exist. Ultimately, we determined an exploratory design using literature analysis would be the most appropriate to facilitate the research objective for this work. More specifically, we determined that an enumerative taxonomy would be appropriate to model the features as measures of effectiveness as well as the relationships between such features.

3.2 Data collection
For data, we collected relevant published literature using primary research indices such as Google Scholar, EBSCOhost, NCLive, and so forth. We used keywords such as *dynamic, honeypot*, *implementation*, and *architecture* in various combinations. Further, we operationally excluded terms such as *API*, *survey*, or *static* in research titles to screen unrelated literature. Overall, we discovered 46 articles in total across our keyword and key phrase searches. These articles encompassed peer reviewed journal papers and conference proceedings. From this superset, we selected a subset of 20 articles based on the presence of enough honeypot design details. These details included technical descriptions, architecture or implementation diagrams, software code or pseudocode, and so forth.

3.3 Limitations and Assumptions
Foremost, this work is limited insofar as the findings are representative of what we were able to derive from existing literature. In other words, it is possible that complementary work exists but has not been published or has been published but is not available. We rationalize this limitation in that common features across honeypots ought to be open to common modes of measurement.

Along those lines, this work is limited by the assumption that common features across various dynamic honeypot technology implies common means of measurement. While the expression of a feature may appear similar, or even identical between honeypots, we recognize variances in underlying programming may lead to variations in measurements. We have observed this, for example, in the way honeypots implement *fingerprinting*. However, we expect that properly tuned measures of effectiveness account for these variations.

4. Results

After analyzing the literature, we found four sets of measures that could be used across various dynamic honeypots in order to quantify their effectiveness in capturing adversarial activity. These four measures are: *fingerprinting*, *data capture*, *deception*, and *intelligence*.

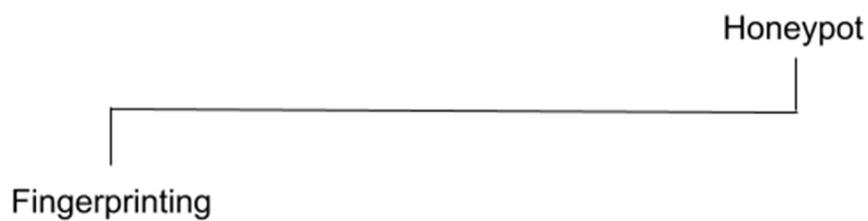

**Fig. 1.** Fingerprinting as the first measure of effectiveness category.

We conceptualize *fingerprinting* as the ability for the honeypot to use a port scanning utility program to identify changes in its own service configuration as well as changes in the service architecture in the environment surrounding the honeypot. Fingerprinting exists as an independent measure, without any connection to the other measures, but is subordinate to the honeypot root (Fig. 1).

There are many different approaches to fingerprinting (Table 1). While available tools such as *nmap* and *xprobe2* are common, fingerprinting could manifest through any utility capable of interrogating services. Effectiveness is therefore measurable in terms of valid, accurate detection of services. Overall, fingerprinting exists as a key component in the taxonomy, as it is a prerequisite for other measures to be effective.

Table 1
Honeypot fingerprinting programs and command input examples

| Noise Level | Programs Used | Command Input |
|---|---|---|
| low | xprobe2 | xprobe2 -r -m 2 -o <file> -X <ip> |
| medium | xprobe2 | xprobe2 -r -m 2 -o <file> -X -T 1-1024,3306 -U 1-1024 <ip> |
| medium-high | nmap | nmap -sT -sU -T3 -sV -O -oX <file> --host-timeout 200 <ip> |
| high | Xprobe2 + nmap | Xprobe2 -r -m 2 -o <file> -X -T 1-1024,3306 -U 1-1024 <ip><br>nmap -sT -sU -T3 -sV -O -oX <file> --host-timeout 200 <ip> |

Note: adapted from Hecker 2012

The second MoE is *data capture*. We conceptualize this as the ability for a honeypot to collect adversarial input with high fidelity. Moreover, we suggest quantification is possible through two subcomponents: capturing the commands an adversary runs in a honeypot and persisting the data beyond the malicious reach of adversaries (Fig. 2).

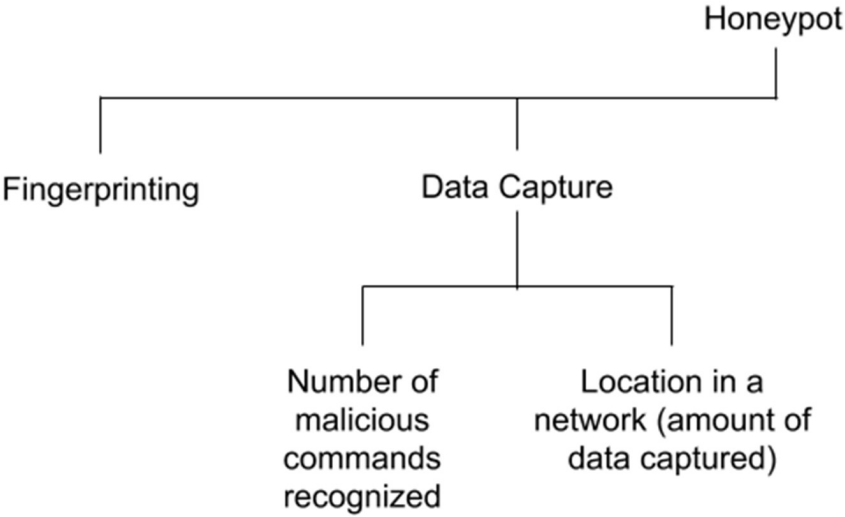

**Fig. 2**. Data capture and subcomponents as the second measure of effectiveness.

While not strictly included in the taxonomy, we suggest categorizing data during persistence according to three types: *events*, *attack*, and *intrusion*. More specifically, an event is any non-harmful command an adversary may run on the system (e.g., *ls*, *cd*). Despite not inherently being malicious, these data may provide correlation points based on user interaction with the system. However, an *attack* constitutes a command that only a malicious user would execute. Building on attack is the most valuable type of command an adversary could run would be an *intrusion*, which we assert is any command that compromises the integrity of the system or interrupts normal, expected system function.

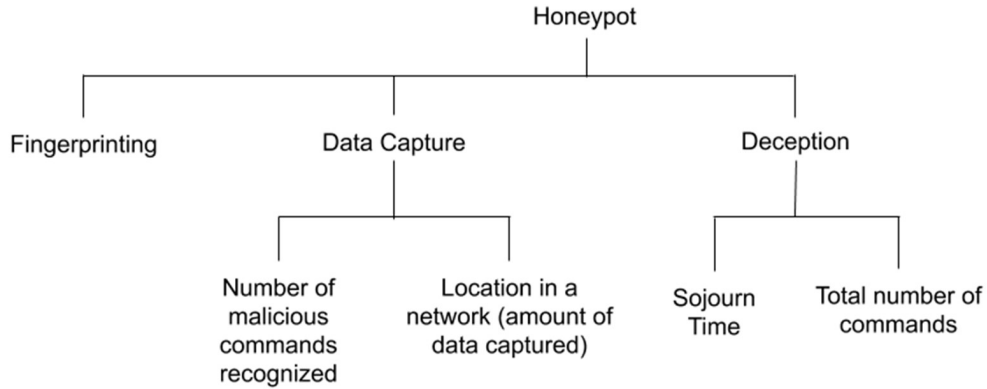

**Fig. 3.** Deception and subcomponents as the third measure of effectiveness.

Deception is the third MoE we extracted from the literature. Broadly, *deception* is how well an adversary is tricked into continued adversarial behavior. This is quantifiable by *sojourn time* or how long an adversary spends in our system in combination with the total number of commands executed (Fig. 3). A longer session time means the honeypot has elicited adversarial behavior effectively. Meanwhile, the other deception subcomponent is the number of commands that an adversary inputs to our honeypot. Not to be confused with data capture where we count the number of commands, here we conceive of total number of commands as a slope indicating the degree of deception. In other words, the more commands entered indicates expanding deceptive effectiveness.

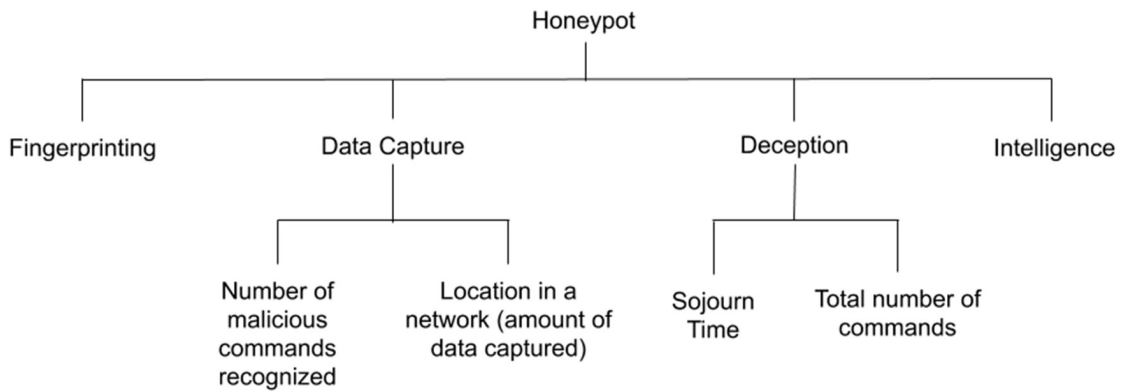

**Fig. 4.** Intelligence as the fourth measure of effectiveness.

The modern era of honeypot research is demarcated by the presence of *intelligence*. Thus, we include intelligence as a MoE inclusive of the machine learning algorithms employed as the means of dynamism in the honeypot. There are a variety of honeypot machine learning implementations both in the context of algorithms as well as features rendered intelligent. For MoE, we specifically intend to measure the degree to which a dynamic honeypot can change and alter itself according to the incoming data capture and deceptive measures (Fig. 4). Ultimately, we organize the MoE according to *levels* (Table 2.).

| Table 2 Levels of artificial intelligence present in honeypots ||
|---|---|
| Intelligence Level | Significance |
| 5 | A Honeypot with a level 5 Intelligence would be able to recognize an adversary in its system and automatically be able to modify the environment around the adversary in order to more effectively deceive them. |
| 4 | A Honeypot at level 4 Intelligence would be able to automatically redeploy itself and harden its defenses after analyzing the logs from previous attacks. This would mean that the honeypot would become more difficult to crack as more people attack it, giving us data into more complex or stronger adversary strategies |
| 3 | An Intelligence level 3 Honeypot can recognize that an adversary is performing malicious commands in a vital area, and can eject the adversary or shut itself off and redeploy before the threat can do any actual damage to the system |
| 2 | A honeypot with level 2 Intelligence can automatically re-deploy itself after it has been shut off without any changes to its initial configuration |
| 1 | A Honeypot with an Intelligence level of 1 is a static honeypot |

5. Conclusions

Honeypots attract adversaries by emulating operating systems, applications, and services with known vulnerabilities [2]. Unfortunately, honeypots are difficult to implement and maintain [4][5]. Dynamic honeypots are especially problematic in this manner [6]. On one hand, existing literature contains a plethora of suggestions as to how dynamic honeypots can be effectively deployed or managed. Yet, on the other hand there is little quantitative validation of effectiveness in this regard which leaves professionals, researchers, and educators without the means to differentiate between implementation or management modalities.

For this reason, we investigated the potential for a standardized set of MoE related to common dynamic honeypot features. Twenty articles of primary research served as the foundation for our analysis. Based on this data set, we selected four quantitatively assessable characteristics of honeypots to measure effectiveness as honeypots and assembled the components into a taxonomy. The proposed model establishes this as *fingerprinting*, *data capture*, *deception*, and *intelligence*. Collectively, these should be successful in quantitatively assessing dynamic honeypots.

5.1 Recommendations
Accordingly, our recommendations and ideas for future work center on exploring this notion of *should be successful* in the context of the measure of effectiveness taxonomy. Based on our findings, we have three recommendations.

First, we recommend the measures of effectiveness be employed as metrics for multiple honeypots currently being used in the industry, such as the Google Hack Honeypot, Honeyd, KFSensor, Cowrie, and so forth. A comparative analysis would provide insight into the effectiveness of honeypot solutions and provide instrument validation.

Along those lines, the taxonomy may translate into a tool that automates data gathering for each of the measures. Furthermore, the tool might generate a score for the subject honeypot as well as for individual services within the honeypot. We recognize the inherent limitation however insofar as such a tool can only measure what the tool can access. Thus, we also suggest that a policy setting forth an appropriate set of honeypot API standards be constructed to aid future efforts.

Another possible avenue of interest would be the use of the MoE model, assessment tool, and API policy, to guide the construction of a more robust honeypot framework. While it is not clear that another honeypot would be of benefit to researchers or practitioners, deeper understanding of adversary behavior and techniques would be beneficial. The new honeypot framework then must emphasize MoE elements.